\documentclass[prc,aps,showpacs,twocolumn,floatfix,superscriptaddress]{revtex4}

\usepackage{epsfig}
\usepackage{graphicx}% Include figure files
\usepackage{dcolumn}% Align table columns on decimal point
\usepackage{bm}% bold math

\begin{document}

\title{Microscopic description of complex nuclear decay: multimodal fission}

\author{A. Staszczak}
\affiliation{Institute of Physics, Maria Curie-Sk{\l}odowska University,
pl.\ M.\ Curie-Sk{\l}odowskiej 1, 20-031 Lublin, Poland}
 \affiliation{Department of Physics and
Astronomy, University of Tennessee Knoxville, TN 37996}
\affiliation{Physics Division, Oak Ridge National Laboratory, P.O.
Box 2008, Oak Ridge, TN 37831}

\author{A. Baran}
\affiliation{Institute of Physics, Maria Curie-Sk{\l}odowska University,
pl.\ M.\ Curie-Sk{\l}odowskiej 1, 20-031 Lublin, Poland}
 \affiliation{Department of Physics and
Astronomy, University of Tennessee Knoxville, TN 37996}
\affiliation{Physics Division, Oak Ridge National Laboratory, P.O.
Box 2008, Oak Ridge, TN 37831}

 \author{J. Dobaczewski}
\affiliation{Institute of
Theoretical Physics, University of Warsaw, ul. Ho\.{z}a 69, 00-681
Warsaw, Poland}
\affiliation{Department of Physics, P.O. Box 35 (YFL),
FI-40014 University of Jyv\"askyl\"a, Finland}

\author{W. Nazarewicz}
 \affiliation{Department of Physics and
Astronomy, University of Tennessee Knoxville, TN 37996}
\affiliation{Physics Division, Oak Ridge National Laboratory, P.O.
Box 2008, Oak Ridge, TN 37831}
\affiliation{Institute of
Theoretical Physics, University of Warsaw, ul. Ho\.{z}a 69, 00-681
Warsaw, Poland}

\date{\today}% It is always \today, today,

\begin{abstract}
Our understanding of nuclear fission, a fundamental nuclear decay,
is still incomplete
due to the complexity of the process. In
this paper, we describe  a study
of spontaneous fission using the symmetry-unrestricted nuclear
density functional theory. Our results show that the observed
bimodal fission can be explained in terms of pathways in
multidimensional collective space corresponding to different
geometries of fission products. We also predict a new phenomenon of
trimodal spontaneous fission for some rutherfordium, seaborgium, and hassium isotopes.
\end{abstract}


\maketitle

\section{Introduction}
Seventy years ago Joliot Curie and Savitch \cite{[Cur38]} discovered
that the exposure of uranium to neutrons leads to the existence of
lanthanum. Following this finding, Hahn and Strassmann
\cite{[Hah39]} proved definitively that bombarding uranium with
neutrons produces alkali earth elements, ushering in what has come
to be known as the atomic age. The term nuclear fission was coined
one year later by Meitner and Frisch \cite{[Mei39a]}, who explained
experimental results in terms of the division of a heavy nucleus
into two lighter nuclei. In 1939, Bohr and Wheeler \cite{[Boh39a]}
developed a theory of fission based on a liquid drop model.
Interestingly, their work also contained an estimate of a lifetime
for fission in the ground state. Soon afterwards, Petrzhak and Flerov
\cite{[Pet40]} presented the first experimental evidence for such
spontaneous fission (SF).

While early descriptions of fission were based on a purely
geometrical framework of the nuclear liquid drop model \cite{[Boh39a]}
(i.e., shape-dependent competition between Coulomb and surface
energy), it was soon realized \cite{[Hil53]} that the single-particle
motion of protons and neutrons moving in a self-deforming mean field
is crucial for the understanding of a range of phenomena such as
fission half-lives, mass and energy distributions of yields, cross
sections, and fission isomers \cite{[Wag91],[Bjo80]}. In the
macroscopic-microscopic method (MMM) proposed by Swiatecki
\cite{[Swi55]} and Strutinsky {\it et al.}  \cite{[Str67a],[Bra72c]},
quantum shell effects are added atop the average (or macroscopic)
behavior described by the liquid drop, and this approach turned out
to be very successful in explaining many features of SF
\cite{[Mol01],[Mol08],[Sob07]}.

Quantum mechanically,
fission represents a time-dependent solution of the many-body
Schr\"odinger equation where all particles move collectively. To fully
solve such a time-dependent problem for more than 200 particles is
neither possible nor sensible because the essence of the process is
in its coherence. Consequently, most of the essential physics should
be contained in underlying mean fields. This determines the choice
of a microscopic tool to be used: the nuclear density functional
theory (DFT) \cite{[Ben03]}. The advantage of DFT is that, while
treating the nucleus as a many-body system of fermions, it provides
an avenue for identifying the essential collective degrees of
freedom.

Because the commonly used nuclear  density functionals are
usually adjusted to nuclear ground-state properties and infinite
nuclear matter, and most applications are symmetry-restricted to
speed up computations, self-consistent theories typically are not as
quantitative as MMM when it comes to SF, except for some cases \cite{[Gou05]}. It is only recently that an
effort has been made to systematically optimize the effective forces
by considering experimental data relevant to large deformations
\cite{[Gor07]}.

In this work, we provide a microscopic description of multi-mode fission
based on the symmetry-unrestricted nuclear DFT. Since many observed
fission characteristics  can be traced back to topologies of fission
pathways in multidimensional collective space \cite{[Mol01]}, allowing
for arbitrary shapes on the way to fission is the key. To this end, we
search for the optimum collective trajectory in a multidimensional
space. The barrier penetration probability, or a fission half-life, is
computed by integrating the action along this optimum path. In practice,
this is done by constraining the nuclear collective coordinates
associated with shape deformations to have prescribed values of the
lowest multipole moments, $Q_{\lambda\mu}$ by which we explore the main
degrees of freedom related to elongation $(\lambda\mu=20)$,
reflection-asymmetry $(\lambda\mu=30)$, and necking $(\lambda\mu=40)$.
The effects due to triaxiality  are known to be important around the top
of the first fission barrier
\cite{[Mol08],[Lar72],[Ben98a],[Bur04],[Sta05],[Del06]}. Indeed,  the
first saddle point is lowered by several MeV by triaxial degrees of
freedom. In our work, we take into account  the impact of non-axial
degrees of freedom   by considering   the triaxial quadrupole moment
($\lambda\mu=22$).

The paper is organized as follows. Section~\ref{model} describes the details of our model. The results of calculations are discussed in Sec.~\ref{calculations}. Finally, Sec.~\ref{conclusions} summarizes the main results of our study and offer perspectives for future work.

\section{The model}\label{model}

The calculations
were carried out using  a symmetry-unrestricted DFT program based on
the Hartree Fock-Bogoliubov solver HFODD \cite{[Dob04c],[Dob09b]} capable
of treating simultaneously all the possible collective degrees of
freedom that might appear on the way to fission. Based on this DFT
framework, we calculated the total energy along the
fission pathways, corresponding
collective inertia (collective masses)
and zero point energy (ZPE) corrections to account for quantum
fluctuations.

In the particle-hole channel, we use the  SkM$^*$ energy density
functional \cite{[Bar82]} that
 has been optimized  at large deformations; hence it is often used
for fission barrier predictions. In the pairing channel, we adopted
 a seniority pairing force with the strength
 parameters fitted to
reproduce the experimental gaps in $^{252}$Fm \cite{[Sta07]}.
Because the nuclei
considered are all well bound, pairing could be treated within the
BCS approximation.
The
single-particle basis consisted of the lowest 1,140 stretched
states originating from the lowest 31 major oscillator shells.

In the analysis of fission pathways, we explored multidimensional
collective space. To separate fission pathways, we computed energy
surfaces in the deformation spaces $\{Q_{20}, Q_{30}\}$ and $\{Q_{20},
Q_{40}\}$. The calculations were not limited to axial shapes; triaxial
deformations appear if energetically favorable (e.g., within the inner
barrier). At each deformation point, defined by the set of constrained
multipole moments, fully self-consistent DFT equations have been solved,
whereupon the total energy of the system is always minimized with
respect to all remaining (i.e., unconstrained) shape parameters. The
optimum 1D paths have been localized in the form of multipole moments,
$Q_{30}$, $Q_{22}$, and $Q_{40}$,  becoming functions of the
driving moment, $Q_{20}$.

The vibrational and rotational ZPE  corrections and the cranking
quadrupole mass parameter were calculated as described in
Ref.~\cite{[Bar07]}. The spontaneous fission half-lives were estimated
from the WKB expression for the double-humped potential barrier
\cite{[Ign69],[Cra70]} assuming a 1D tunneling path along $Q_{20}$.

\section{Results}\label{calculations}

To demonstrate the validity and generality of our method, we chose a
case where several fission pathways were known to coexist and all
intrinsic symmetries of the nuclear mean field were broken. In this
respect, a phenomenon known as bimodal fission, observed in several
fermium and transfermium nuclei \cite{[Bri84],[Hul86a],[Hul89],[Hof89b]},
is a perfect testing ground. It manifests itself, for example, in a
sharp transition from an asymmetric mass division in $^{256}$Fm and
$^{256}$No to a symmetric mass split in $^{258}$Fm and $^{258}$No.
Furthermore, the total kinetic energy distributions of the
fission fragments appear to have two peaks centered around 200 and
233 MeV. It has been suggested
\cite{[Hul89],[Bro86a],[Mol87a],[Pas88a],[Cwi89d],[Zha00]} that the higher
energy fission mode corresponds to a scission configuration
associated with two touching, nearly spherical, fragments with the
maximal Coulomb repulsion, whereas the lower-energy mode can be
associated with more elongated fragments.
Before this work, bimodal fission was studied within the MMM
\cite{[Bro86a],[Mol87a],[Pas88a],[Cwi89d],[Mol08]} and nuclear DFT
\cite{[Ben98a],[Del06],[War02],[War06],[Bon06],[Dub08]}. All those studies
were symmetry-restricted (i.e., they did not consider simultaneous
inclusion of elongation, triaxiality, and reflection-asymmetry).

\subsection{Fission pathways of the fermium isotopes}
\begin{figure}[htb]
 \centerline{\includegraphics[trim=0cm 0cm 0cm
0cm,width=0.45\textwidth,clip]{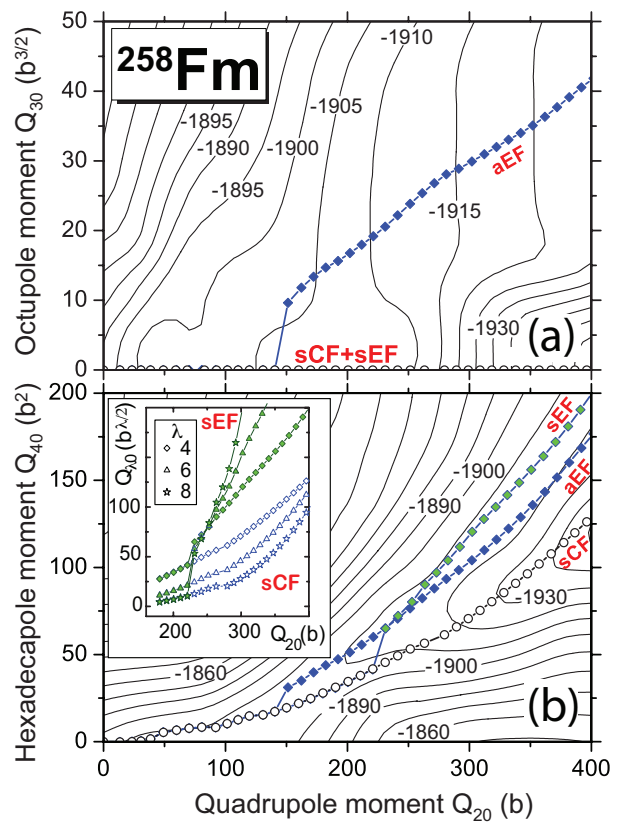}}
\caption{(Color online) Two-dimensional total energy surfaces for $^{258}$Fm in the
plane of  collective coordinates: $Q_{20}$-$Q_{30}$ (a), and
 $Q_{20}$-$Q_{40}$ (b).  The fission
pathways are marked: symmetric compact fragments (sCF),
symmetric elongated fragment (sEF),
and asymmetric
elongated fragments (aEF) pathways.
The difference between contour lines is 5\,MeV in (a) and 10\,MeV in (b).
The asymmetric trajectory aEF bifurcates away near $Q_{20}\approx
150$\,b from $Q_{30}$=0 while
the bifurcation between sEF and sCF occurs near
$Q_{20}\approx 225$\,b.
The inset shows the multipole moments $Q_{\lambda 0}$ along sCF and sEF.
}
\label{fig1}
\end{figure}
To identify saddle points and fission pathways
in a multidimensional energy surface is
not a simple task. As pointed out in earlier studies
\cite{[Mol01],[Mol08],[Mye96],[Mam98]}, saddle points obtained in calculations
constrained by only one collective variable are sometimes incorrect;
hence, special numerical techniques are required to find them. To
this end,  we computed
two-dimensional (2D) energy surfaces in $\{Q_{20}, Q_{30}\}$ and
$\{Q_{20}, Q_{40}\}$ planes.
Based on the initial 2D uniform  grids, $Q_{20}$=0(10)400\,b,
$Q_{30}$=0(5)50\,b$^{3/2}$, and $Q_{40}$=0(10)200\,b$^2$,
we calculated the constrained HF+BCS energy.
In our HFODD calculations we employed the standard method of quadratic  constraint  (the quadratic penalty method) \cite{[Gir70],[Flo73]}.
Within this approach, the moments calculated from the converged densities differ slightly from the requested values  defining the constraints \cite{[Flo73]}. This implies that the final mesh used for interpolation is non-uniform.
Using the self-consistent values of multipole moments,
the non-uniform  interpolation was carried out to produce the
final  result.
Figure~\ref{fig1} shows 2D energy surfaces for
$^{258}$Fm.

The 1D fission pathways shown in Fig.~\ref{fig1} were obtained by
finding the local energy minimum at a given value of $Q_{20}$ and
following this solution when gradually increasing the driving moment  $Q_{20}$. This method
thus relies on local properties of the $Q_{20}$-constrained energy
surface. In practice, one obtains different energy surfaces locally
valid around each fission pathway. In order to facilitate the
presentation, however, we carried out interpolation between different
energy sheets. For that reason,   the contour lines  in
Fig.~\ref{fig1}  slightly depend on the  algorithm used to
interpolate between 2D constrained results; hence, the 2D surfaces
should be considered as a qualitative guidance on the topological
structure of the energy surface.

Beyond the first barrier, at $Q_{20}\approx
150$\,b, a reflection-asymmetric path corresponding to asymmetric
elongated fragments (aEFs) branches away from the symmetric valley,
see Fig.~\ref{fig1}a.
At $Q_{20}\simeq 225$\,b, a reflection-symmetric path splits into
two branches: one corresponding to nearly spherical symmetric
compact fragments (sCFs) and one associated with symmetric elongated
fragments (sEFs). This bifurcation is clearly seen in Fig.~\ref{fig1}b.
Such three fission pathways
were predicted in early work based on MMM \cite{[Bro86a],[Pas88a]} and
also found recently within a DFT framework \cite{[War06],[Bon06]},
except that the axial-symmetry was enforced in all these studies.

\begin{figure}[htb]
 \centerline{\includegraphics[trim=0cm 0cm 0cm 0cm,width=0.48\textwidth,clip]{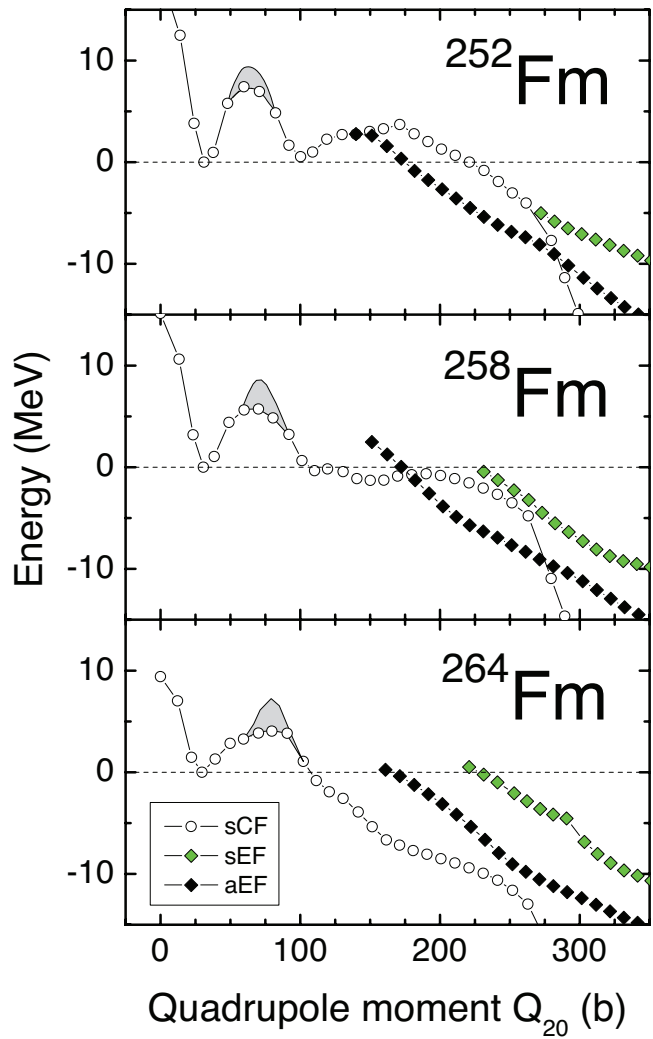}}
\caption{(Color online)
Calculated 1D fission pathways for $^{252,258,264}$Fm as functions of the driving quadrupole moment, $Q_{20}$. Open circles, light diamonds, and dark diamonds  denote the symmetric compact fragment (sCF), symmetric elongated fragment (sEF), and asymmetric elongated fragment (aEF) valleys, respectively.
The effect of triaxiality is important in the region of the first barrier; the corresponding energy gain  is marked by gray shading.
}
\label{fig1b}
\end{figure}

A pattern of
similarly competing fission valleys was found for all investigated
 isotopes. As an illustrative example, the results of our   1D fission pathway calculations for $^{252,258,264}$Fm
are displayed in Fig.~\ref{fig1b}.
(The energy curves corresponding to individual pathways in  $^{256,260}$Fm can be found in our earlier work \cite{[Sta07]}.)
The transition from an asymmetric fission path
in $^{252}$Fm to a compact-symmetric path in $^{264}$Fm is due to shell effects in
the emerging fission fragments approaching
the doubly magic nucleus $^{132}$Sn \cite{[Hul89],[Mol87a],[Poe87],[Bha91],[Gon93]}.
The triaxial deformations are important around the
first (inner) fission barrier, and they reduce the fission barrier
height by several MeV.

While the asymmetric pathway aEF is favored in the lighter Fm
isotopes, e.g.,   $^{252}$Fm,
both symmetric paths are open for $^{258}$Fm, due to the
disappearance of the outer fission barrier in sCF and sEF. It is to be noted that the symmetric pathways sCF and sEF in $^{258}$Fm are predicted to
bifurcate away well outside the first barrier (see also  recent
work \cite{[Ich09]} based on MMM).

In the case of $^{260-264}$Fm, we find that
there is no outer potential barrier along the sCF trajectory, and
the sEF and aEF paths lie significantly higher in the outer region.
It should be emphasized that the pathways correspond to different regions
of the collective space and this is apparent when
studying them  in more than one dimension. Indeed, aEF is
well separated from sCF and sEF  in $Q_{30}$ (the apparent crossing between aEF and sEF in Fig.~\ref{fig1}b  is
an artifact of the 2D projection) while the symmetric
trajectories sCF and sEF strongly differ in the values of higher multipole
moments $\lambda$=4, 6, and 8 (see the inset in Fig.~\ref{fig1}).

\subsection{Spontaneous fission half-lives}

 \begin{figure}[htb]
 \centerline{\includegraphics[trim=0cm 0cm 0cm
0cm,width=0.40\textwidth,clip]{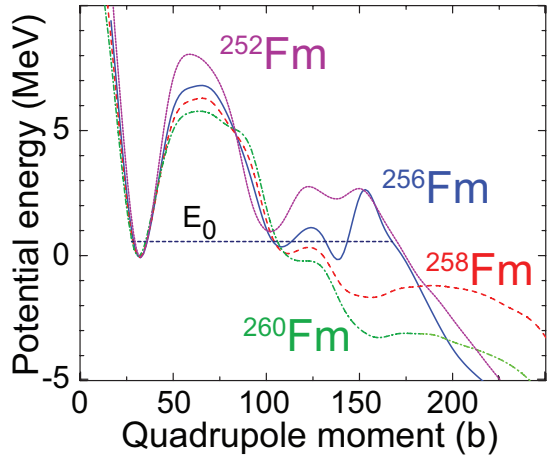}}
\caption{(Color online) Potential
energy curves of $^{252}$Fm, $^{256}$Fm, $^{258}$Fm, and $^{260}$Fm along
optimum 1D paths containing the quantum zero-point energy correction,
drawn in the common scale relative to values calculated at the
ground-state minima.}
\label{fig2}
\end{figure}
Having determined the lowest fission valleys, i.e.,
the lowest energy pathways along $Q_{20}$,
we computed the corresponding
ZPE corrections. The resulting collective potentials are plotted in
Fig.~\ref{fig2} for $^{252}$Fm, $^{256}$Fm, $^{258}$Fm, and
$^{260}$Fm.
It is seen that when going from $^{252}$Fm to
$^{260}$Fm, the first barrier gets reduced and the outer barrier
disappears altogether.

\begin{figure}[htb]
 \centerline{\includegraphics[trim=0cm 0cm 0cm
0cm,width=0.40\textwidth,clip]{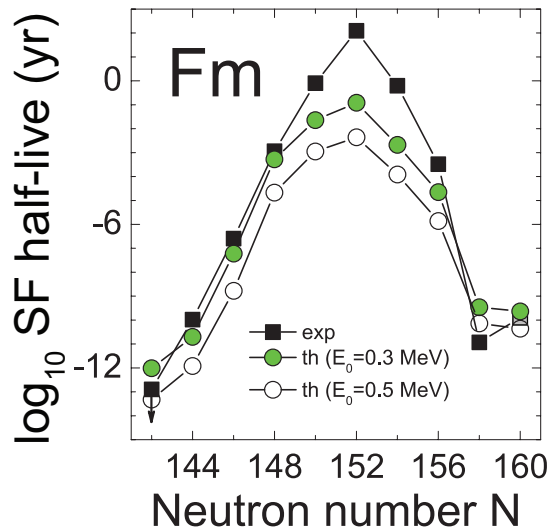}}
\caption{(Color online)  Fission half lives of even-even fermium
isotopes with 242$\leq$A$\leq$260, calculated in this study
compared with experimental data \cite{[Hol00a],[Khu08]} for the
two values of the zero-point energy $E_0$=0.3\,MeV and 0.5\,MeV.
}
\label{fig3}
\end{figure}

To assess the SF half-lives theoretically, we
calculated the collective inertia parameter along $Q_{20}$ and
performed WKB barrier penetration calculations
for even-even fermium isotopes with 242$\leq$A$\leq$260.
We assumed two values of the ground state energy counted
from the ground state potential energy minimum:
$E_0$=0.3\,MeV and the commonly used  value \cite{[Nil69]} of 0.5\,MeV. The
resulting SF half-lives are shown in Fig.~\ref{fig3}.
In spite of a
fairly simple 1D penetration picture, it is satisfying to see a
quantitative agreement between experiment \cite{[Hol00a],[Khu08]} and theory
(for Gogny-DFT results, see \cite{[War06]}). The existence of a small
outer barrier in $^{256}$Fm is significant as it increases the
fission half-life in this nucleus by more than four orders of
magnitude compared to that of $^{258}$Fm, thus explaining the rapid
change in experimental SF half-lives between these nuclei.

\subsection{Multimodal fission}

To map out the competition between different fission pathways in the heaviest elements, we carried out systematic calculations for even-even nuclei with 98$\le$$Z$$\le$108 and 154$\le$$N$$\le$160.
A transition from the usual asymmetric fission
channel seen in the actinides to compact symmetric fission is seen
when moving towards $^{264}$Fm. In the intermediate region of
bimodal fission, two symmetric channels coexist. Around $^{260}$Sg
($Z$=106, $N$=154), our calculations predict trimodal fission, i.e., competition between the asymmetric fission valley and two
symmetric ones. (The term ``multimodal fission"
has been previously used by M.G. Itkis {\it et al.} \cite{Itkis} in the context of fusion-fission and  quasi-fission of  hot superheavy nuclei produced in heavy ion reactions.)
\begin{figure}[htb]
 \centerline{\includegraphics[trim=0cm 0cm 0cm
0cm,width=0.48\textwidth,clip]{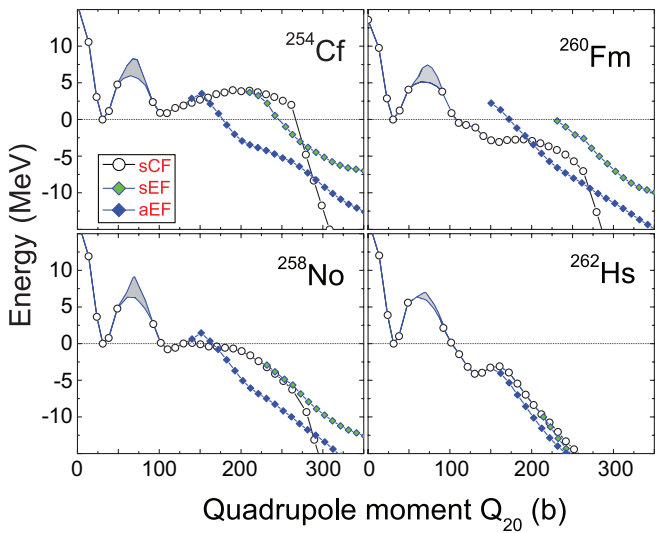}}
\caption{(Color online) Similar as in Fig.~\ref{fig1b} except for fission pathways in
$^{254}$Cf (asymmetric fission),
$^{260}$Fm (symmetric compact fission), $^{258}$No (bimodal fission)
and $^{262}$Hs (trimodal fission).
}
\label{fig4}
\end{figure}

The representative fission pathways for $^{254}$Cf (asymmetric fission),
$^{260}$Fm (symmetric compact fission), $^{258}$No (bimodal fission)
and $^{262}$Hs (trimodal fission) are displayed in Fig.~\ref{fig4}.
The inclusion of triaxiality  significantly reduces the inner
barrier in  $^{254}$Cf,
$^{260}$Fm, and  $^{258}$No while the effect in $^{262}$Hs is much weaker.
\begin{figure}[htb]
 \centerline{\includegraphics[trim=0cm 0cm 0cm
0cm,width=0.45\textwidth,clip]{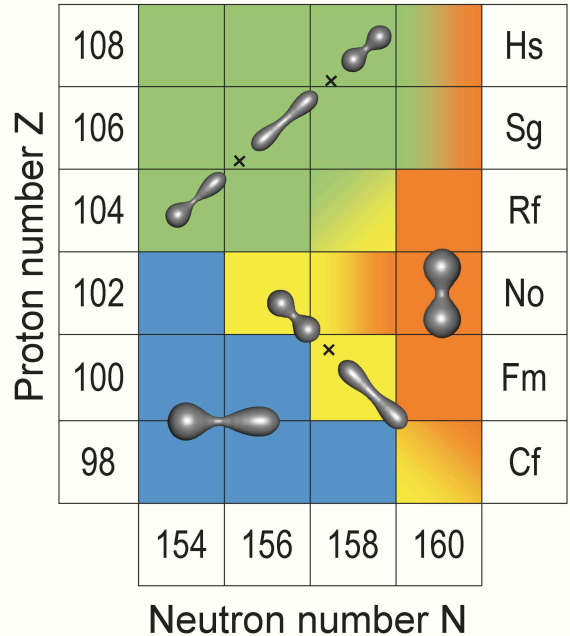}}
\caption{(Color online) Summary of fission pathway results obtained in the present
study. Nuclei around $^{252}$Cf$_{154}$ are
predicted to fission along
the asymmetric path aEF; those around $^{262}$No$_{160}$ along the symmetric
pathway sCF. These two regions are separated by the
bimodal
symmetric fission (sCF + sEF) around $^{258}$Fm$_{158}$. In a number of
the Rf, Sg, and Hs nuclei, all three fission modes
are likely (aEF + sCF + sEF; trimodal fission).
In some cases, labelled by two-tone shading with one tone
dominant, calculations predict coexistence of two decay scenarios
with a preference for one. Typical
nuclear shapes
corresponding to the calculated nucleon densities are marked.}
\label{fig5}
\end{figure}
Other nuclei from this region exhibit similar pattern of fission pathway competition. A  summary of our  findings is given in
Fig.~\ref{fig5}.

\section{Summary}\label{conclusions}

In summary, the symmetry-unrestricted nuclear DFT framework has been applied to
the problem of SF. As an example, we studied the challenging case of
static SF pathways in $^{256}$Fm, $^{258}$Fm, and $^{260}$Fm and in
a number of neighboring nuclei. We found competition between
symmetric-compact, symmetric-elongated, and asymmetric-elongated
fission valleys that is consistent with the observed distribution of
fission yields. The saddle points obtained in constrained 1D
calculations were confirmed through an analysis of 2D energy
surfaces. From the calculated collective potential and collective
mass, we estimated SF half-lives, and good agreement with
experimental data was found. Finally, we predicted trimodal fission
for several rutherfordium, seaborgium, and hassium isotopes.

It is worth noting that calculations of self-consistent energy 2D
surfaces are computer intensive. Because a single HFODD run with all
self-consistent symmetries broken takes about 60 minutes of CPU
time, it takes about 3 CPU-years to carry out the full fission
pathway analysis for 24 nuclei; hence, massively parallel computer
platforms had to be used.

In the near
term, we intend to improve the theory of SF half-lives by
considering multidimensional inertia tensors and by performing the
direct minimization of the collective action in a multidimensional
collective space \cite{[Bar81]}. In the long term, the theory will be
extended to account for nonadiabatic effects (e.g., along the lines
of Refs.~\cite{[Neg89a],[Ska08]}). In addition, quality microscopic input for
fission calculations is needed. Of particular importance is the
development of the nuclear energy density functional better
reproducing both bulk nuclear properties and spectroscopic data.

This work was supported in part by the National Nuclear Security
Administration under the Stewardship Science Academic Alliances
program through U.S. Department of Energy Research Grant
DE-FG03-03NA00083; by the U.S. Department of Energy under Contract
Nos. DE-FG02-96ER40963 (University of Tennessee), DE-AC05-00OR22725
with UT-Battelle, LLC (Oak Ridge National Laboratory), and
DE-FC02-09ER41583 (UNEDF SciDAC Collaboration); by the Polish
Ministry of Science and Higher Education under Contract
No. N~N~202~328234; and by the Academy of
Finland and University of Jyv\"{a}skyl\"{a} within the FIDIPRO
program. Computational resources were provided by the National
Center for Computational Sciences at Oak Ridge National Laboratory.

%\bibliographystyle{unsrt}
%\bibliography{C:/Actual/LaTeX/Latex.all/jacwit25,bimodal}

\end{document}